\nonstopmode
\pdfoutput=1

\documentclass[preprint, amsmath, floatfix, aps, superscriptaddress]{revtex4}
\usepackage{graphicx}
\usepackage{hyperref}
\usepackage[utf8x]{inputenc}
\usepackage[T1]{fontenc} 
\hypersetup{pdfauthor={Matthias Kammerer},pdftitle={Magnetic Vortex Core Reversal by Excitation of Spin Waves}}
\usepackage{natbib} 

\newcommand{\x}{\,$\times$\,}

\begin{document}
\title{Magnetic Vortex Core Reversal by Excitation of Spin Waves}
\author{Matthias Kammerer}
\email[email: ]{kammerer@mf.mpg.de}
\author{Markus Weigand}
\author{Michael Curcic}
\author{Matthias Noske}
\author{Markus Sproll}
\affiliation{Max-Planck-Institut für Metallforschung, Heisenbergstraße 3, 70569 Stuttgart, Germany}
\author{Arne Vansteenkiste}
\author{Bartel Van~Waeyenberge}
\affiliation{Department of Solid State Sciences, Ghent University, Krijgslaan 281 S1, 9000 Ghent, Belgium}
\author{Hermann Stoll}
\affiliation{Max-Planck-Institut für Metallforschung, Heisenbergstraße 3, 70569 Stuttgart, Germany}
\author{Georg Woltersdorf}
\author{Christian H. Back}
\affiliation{Institut für Experimentelle und Angewandte Physik, Universität Regensburg, Universitätsstraße 31, 93053 Regensburg, Germany}
\author{Gisela Schuetz}
\affiliation{Max-Planck-Institut für Metallforschung, Heisenbergstraße 3, 70569 Stuttgart, Germany}
\date{\today}

\begin{abstract}

Micron-sized magnetic platelets in the flux closed vortex state are characterized by an in-plane curling magnetization and a nanometer-sized perpendicularly magnetized vortex core. Having the simplest non-trivial configuration, these objects are of general interest to micromagnetics and may offer new routes for spintronics applications. Essential progress in the understanding of nonlinear vortex dynamics was achieved when low-field core toggling by excitation of the gyrotropic eigenmode at sub-GHz frequencies was established. At frequencies more than an order of magnitude higher vortex state structures possess spin wave eigenmodes arising from the magneto-static interaction.
Here we demonstrate experimentally that the unidirectional vortex core reversal process also occurs when such azimuthal modes are excited. These results are confirmed by micromagnetic simulations which clearly show the selection rules for this novel reversal mechanism. Our analysis reveals that for spin wave excitation the concept of a critical velocity as the switching condition has to be modified.

\end{abstract}

\maketitle

Thin film soft magnetic platelets of suited micron or submicron size are characterized by an in-plane closed flux magnetization, minimizing the dipolar energy. However, at the centre, the exchange energy forces the magnetization out-of-plane in a small area of only a few exchange lengths in diameter \cite{Shinjo2000, Wachowiak2002} creating the vortex core with a distinct polarization $p$, either up ($p = +1$) or down ($p = -1$). The vortex introduces an important sub-GHz mode to the excitation spectrum of the magnetization \cite{Huber1982}, which corresponds to the gyrotropic motion of the vortex.

Recently, dynamic vortex core reversal was discovered by application of pulsed magnetic fields \cite{Xiao2006, Hertel2007, Weigand2009}, alternating magnetic fields \cite {VanWaeyenberge2006, Curcic2008, Lee2008, deLoubens2009} or spin polarized currents \cite{Yamada2007,Yamada2008}. These discoveries did not only open up the possibility of using magnetic vortices as memory bits and further spintronics applications \cite{Kim2008, Bohlens2008, Pigeau2010}, but also initiated wide investigation of the physics behind this reversal mechanism.

The reversal process appeared to be universal and independent of the type of excitation: it proceeds through the creation and subsequent annihilation of a vortex-antivortex pair \cite{VanWaeyenberge2006, Hertel2007, Guslienko2008100}. The source of this pair is the dynamical distortion of the otherwise stable vortex: it forms a 'dip' -- an oppositely polarized region close to the moving vortex core \cite{Volkel1994,Vansteenkiste2009}. Guslienko {\it et al.} \cite{Guslienko2008100} explained this 'dip'-formation by an effective field, the gyrofield, which is proportional to the vortex core velocity. This leads to the concept of a critical velocity for switching, which is about 320 m/s in Permalloy \cite{Guslienko2008100}.

Furthermore, the sense of rotation of a freely gyrating vortex is determined by the vortex core polarization \cite{Thiele1973,Huber1982,Choe2004}, a property which manifests itself in a high asymmetry of the gyrotropic resonance excited by rotating magnetic fields with different rotation senses \cite{Lee2008}. This can be utilized to selectively reverse the core polarization, as shown experimentally by Curcic {\it et al.} \cite{Curcic2008}. Micromagnetic simulations from Kravchuk {\it et al.} \cite{Kravchuk2007,Kravchuk2009} have indicated that rotating magnetic fields can even induce vortex core reversal at GHz frequencies as a result of the vortex spin wave interaction.

Here we present experiments which verify that polarisation selective vortex core reversal can indeed be achieved by spin wave excitation with in-plane rotating GHz magnetic fields. However, compared to the results of Kravchuk et al. \cite{Kravchuk2007} where only one spin wave mode was consiered, we will present selection rules for different modes. And in contrast to the calculations given by Kravchuk et al. \cite{Kravchuk2009} where the vortex core was artificially fixed during spin wave excitation, we will demonstrate by micromagnetic simulations that the core motion is essential and explains the differences observed in the vortex core switching triggered by distinct spin waves.

\section*{Results}

The presence of the vortex core does not only introduce the gyrotropic mode, but also modifies the high-frequency spin wave normal mode spectrum \cite{Park2003,Buess2005, Buess2004, neudecker2006} by lifting the degeneracy of the azimuthal modes with opposite rotation sense \cite{Zaspel2005, Park2005, Zhu2005, Hoffmann2007, Guslienko2008101, Guslienko2010}.
The out-of-plane component of the magnetization $\Delta m_z$ of these azimuthal spin waves can be described in polar coordinates $(\rho,\phi)$ as \cite{Buess2005} :
\begin{equation}
               \Delta  m_{z,n,m}(\rho, \phi,t)=  r_{z,n} (\rho) e^{i m \phi} e^{- i \omega_{n,m,p} t}
\end{equation}
These modes are characterised by a radial mode number $n$ (positive integer values starting with 1)
denoting the number of nodes of the radial part within the disk and an azimuthal mode number $m$ (integer values), with $\lvert 2 m \rvert$ being the number of nodes in the azimuthal part. The sign of $m$ denotes the rotation sense of the azimuthal wave (CCW: $m$ positive , CW: $m$ negative). The symmetry in the frequency splitting of the azimuthal mode can be written as $\omega_{n,m,-p} =\omega_{n,-m,p}$.

\begin{figure}
\centerline{\includegraphics[clip, width=0.5\textwidth]{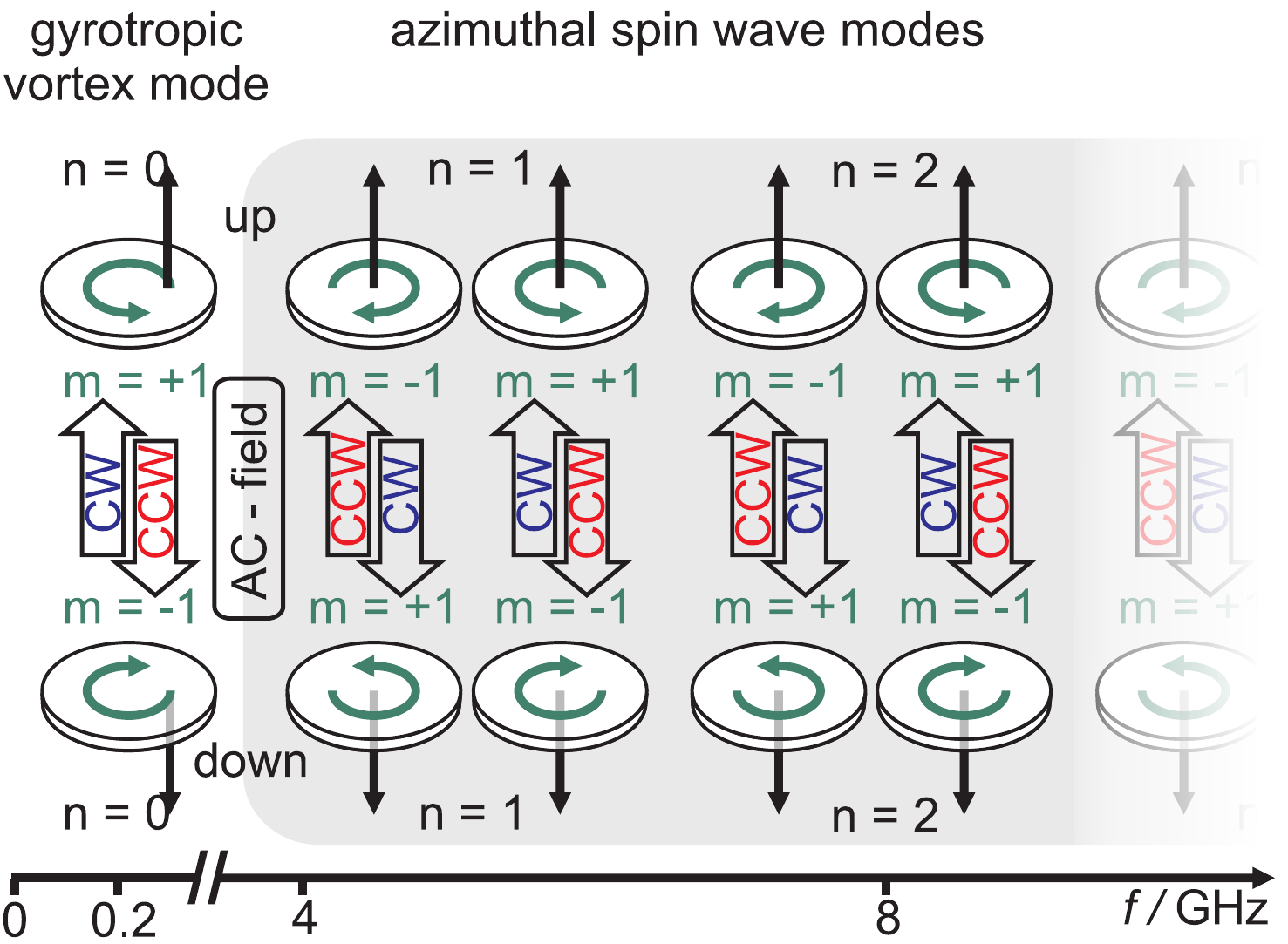}}
\caption{
{\bf Illustration of unidirectional vortex core reversal by external in-plane rotating magnetic fields.} Switching only occurs if the senses of rotation (CW or CCW) of both the external field and the eigenmode (green arrows) are the same. At the left hand side, the sub-GHz frequency gyromode is illustrated. The right hand side shows the azimuthal spin wave modes at much higher (GHz) frequencies, characterized by the radial mode number $n$ and the azimuthal mode number ($m=\pm 1$), denoting the sense of rotation of the eigenmode. In vortex structures the symmetry is broken by the out-of-plane component of the core and thus a frequency splitting is observed between ($m = -1$) and ($m = +1$) modes.
}
\label{Fig1}
\end{figure}

A general concept of resonant vortex core reversal by rotating magnetic fields is sketched in Fig.~\ref{Fig1}. On the left-hand side, the gyrotropic mode ($n=0$, $m=1$), is shown with a CCW sense of rotation for a core up state, and a CW sense of rotation for a core down state. The vortex core can only be reversed resonantly by applying an external rotating magnetic field if its frequency and sense of rotation correspond to the mode, as already shown experimentally in \cite{Curcic2008}. On the right hand side of Fig.~\ref{Fig1}, the frequency-split azimuthal modes are shown. For a vortex up this means that the mode with a CW sense of rotation, denoted by ($n$, $m=-1$), has a lower frequency and the mode with a CCW sense of rotation, denoted by ($n$, $m = +1$), has a higher frequency. For a vortex core down the sign of the azimuthal mode number $m$, indicating the sense of rotation, is reversed for the higher and lower frequency modes compared to vortex core up, as sketched at the bottom part of Fig.~\ref{Fig1}. These spin wave modes with ($n$, $m = \pm 1$) can be selectively excited by the application of rotating in-plane magnetic fields with the corresponding frequency and rotation sense. Thus at a fixed GHz frequency and core polarity the excitation is polarization selective in the same way as for the sub-GHz gyrotropic mode: only rotating fields with one rotation sense will pump the system. This suggests that a similar polarization selective switching mechanism for the vortex core may exist when any of the azimuthal spinwave modes are excited, albeit at much higher frequencies.

\subsection*{Experimental Results}

We excited ferromagnetic Permalloy discs, 1.6\,$\mu$m in diameter, 50\,nm thick, in the vortex magnetic ground state with bursts of GHz rotating magnetic fields of variable frequency and amplitude. The fields were applied with either a clockwise (CW) or counterclockwise (CCW) rotation sense \cite{Curcic2008}. Magnetic scanning soft X-ray microscopy at the MAXYMUS endstation, Helmholtz Zentrum Berlin, BESSY II was used to determine the out-of-plane magnetization of the vortex core.                   

\begin{figure}
\centerline{\includegraphics[width=0.4\textwidth]{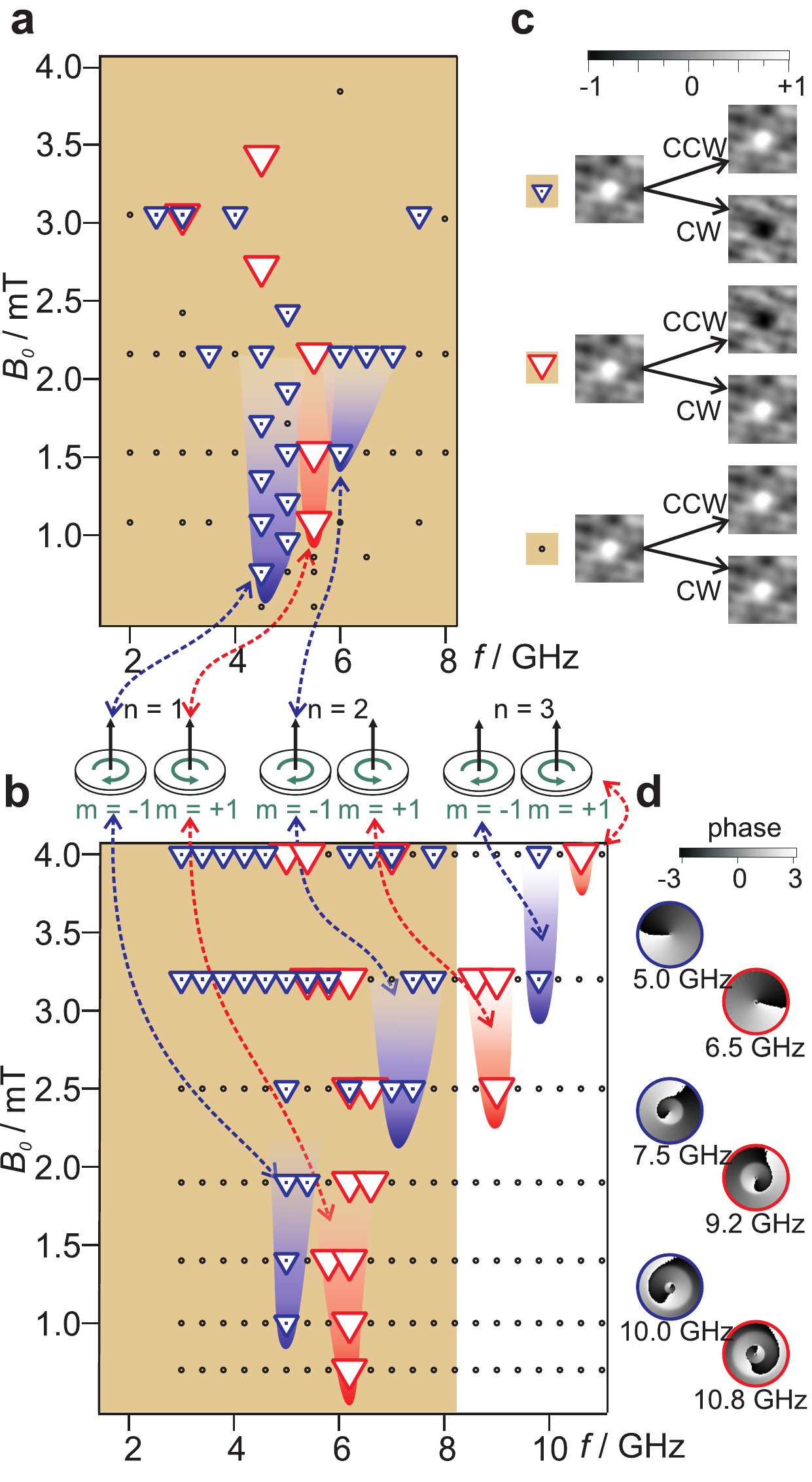}}
\caption {
{\bf Switching phase diagrams.} The legend (c) shows microscopy images of X-ray transmission through the inner part (150 nm times 150 nm) of the sample indicating the vortex core polarity (white: vortex up; black: vortex down) before and after a field burst. The phase diagrams show the points (excitation amplitude vs. frequency) where vortex core reversal from up to down was observed in the experiments (a) and the simulations (b). Rotating in-plane magnetic field bursts with an amplitude $B_0$, a frequency $f$ and a duration of 24 periods have been applied. Like indicated in the legend of the top panel with the recorded X-ray images (c), the blue triangles with a dot in the middle indicate vortex core switching only after a CW rotating field burst, while red triangles indicate switching only after a CCW field burst. Black dots indicate no switching for either rotation sense. The minima in the switching threshold with differing sense of rotation correspond to the resonance frequencies of the excited azimuthal spin wave modes with the same sense of rotation as sketched in the middle part of Fig.~\ref{Fig2}. The modes are identified with the help of the phases derived from a local Fast Fourier Transform of the simulated out-of-plane magnetization of the sample with $1.62 \mu m$ in diameter as shown in the inset to the right of the bottom panel (d).
}
\label{Fig2}
\end{figure}

The experimental verification of the concept as described before is presented in Fig.~\ref{Fig2} (upper part). It shows the vortex core switching events from the initial vortex core up state to the down state as a function of excitation frequency, amplitude $B_0$ and sense of rotation of the external field.  Here, the burst length was fixed at 24 periods of the excitation frequency. Three regions with a minimum in amplitude for vortex core reversal can be observed in the experimental data. Dependent on the sense of rotation of the external magnetic field they are marked with blue or red backgrounds. The regions for CW excitation at about 4.5\,GHz and 6.5\,GHz, CCW excitation at about 5.5\,GHz can be assigned to the ($n=1$, $m=-1$), ($n=2$, $m=-1$) and ($n=1$, $m=+1$) spin wave modes,  respectively (cf. Fig.~\ref{Fig1} and the middle part of Fig.~\ref{Fig2}).

\subsection*{Micromagnetic Simulations}
In order to further investigate the core reversal at these different modes, we performed micromagnetic simulations \cite{OOMMF} for the same sample size and experimental conditions. The results are shown in Fig.~\ref{Fig2}, lower part and are in good agreement with our experimental results. The small differences between the experimental and simulated results can be attributed to the uncertainty in the thickness, saturation magnetization and damping coefficient of the material, as well as to possible structural imperfections in the samples. The phases of a local Fourier analysis \cite{Park2003, Buess2004} of the dynamic magnetization let us clearly identify the azimuthal modes (cf. Fig.~\ref{Fig2}, lower right part). The simulations with an extended frequency range up to 10.5\,GHz reveal even more spin wave eigenmodes capable of vortex core reversal: the ($n=2$, $m=+1$) mode at about 9\,GHz and the two ($n=3$) spin wave modes at about 10 and 10.3\,GHz.

Both the experiments and the micromagnetic simulations show that resonant switching of the vortex core from the down state back to the up state at the same frequency can be achieved only if the sense of rotation of the external field is reversed. This allows selective and unidirectional switching at each resonance frequency.

\begin{figure}
\centerline{\includegraphics[clip, width=0.5\textwidth]{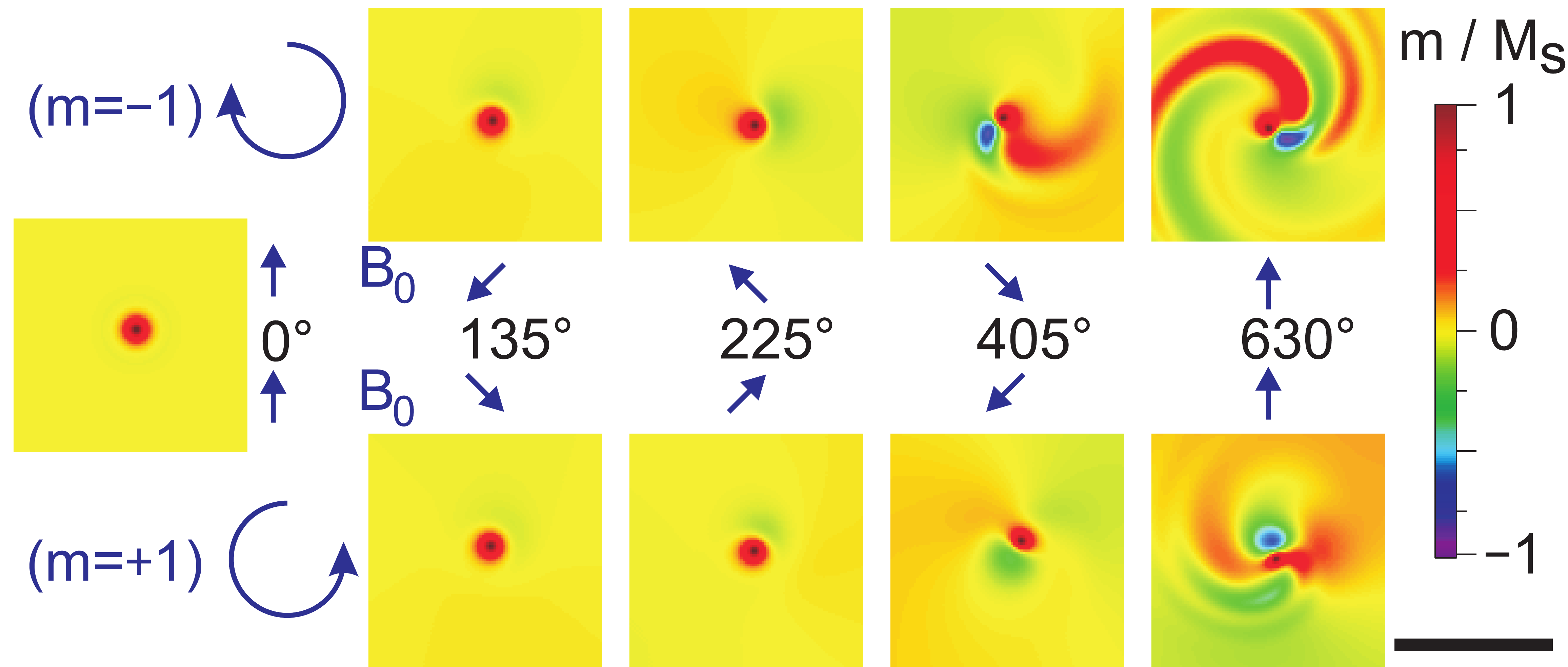}}
\caption{
{\bf Snapshots during spin wave excitation before vortex core reversal.} The frames show the time evolution of the out of plane magnetization for a vortex up during the application of in plane rotating magnetic fields. Only the inner part of the sample is shown. The size of the black bar corresponds to a length of 200 nm. The left frame corresponds to the relaxed ground state (phase angle 0°). The two rows oppose counter rotating modes at a frequency of 5.0 GHz for the ($m=-1$) mode and 6.2 GHz for the ($m=+1$) mode at the same azimuthal angle of the external field. The blue arrows in the middle indicate this angle for the corresponding frame.
}
\label{Fig7}
\end{figure}

Detailed analysis of the micromagnetic simulations reveal that the excitation of all spinwave modes also leads to the creation of a 'dip' near the vortex core, resulting in the same core reversal mechanism as found for gyrotropic excitation \cite{VanWaeyenberge2006, Hertel2007, Vansteenkiste2009}. It is based on the creation and annihilation of a vortex-antivortex pair. But the 'dip' creation itself differs dependent on which spin wave mode is excited.
This 'dip' creation for a vortex up ($p=+1$) is illustrated by snapshots of the simulations in Fig.~\ref{Fig7} for the modes ($n=1$, $m=\pm1$). In the first frame (0°), the unperturbed out-of-plane magnetization of the vortex can be seen. After a short time of excitation (135° phase angle) the azimuthal spin wave with the same rotation sense as the external field develops. The interaction of the spin waves with the gyrotropic mode leads to a displacement of the vortex core towards the oppositely polarized region of the bipolar mode, resulting in a gyration with the same frequency as the external field (225° phase angle).  At 405°, a negatively polarized out-of-plane region can be observed next to the core, to the outside with respect to the centre. It is gyrating around the centre almost in phase with the vortex core. In the ($m=-1$) case (upper panels), this region develops into a 'dip' which finally leads to core reversal. But surprisingly, for the ($m=+1$) in the lower panels, there is an additional negatively polarized region. This region is situated towards the structure centre where the azimuthal mode has a positive polarization and gyrates with a phase shift of about 180° with respect to the vortex core. As can be seen from the last frame at 630°, it is this region which evolves into a fully out-of-plane 'dip', creating the vortex-antivortex pair for switching. The typical magnetization configurations  are shown in Fig.~\ref{Fig3} including the higher radial mode numbers. The trajectories for the vortex core as well as for the 'dips' and their relative phases are illustrated by arrows.

\section*{Discussion}
\begin{figure}
\centerline{\includegraphics[clip, width=0.5\textwidth]{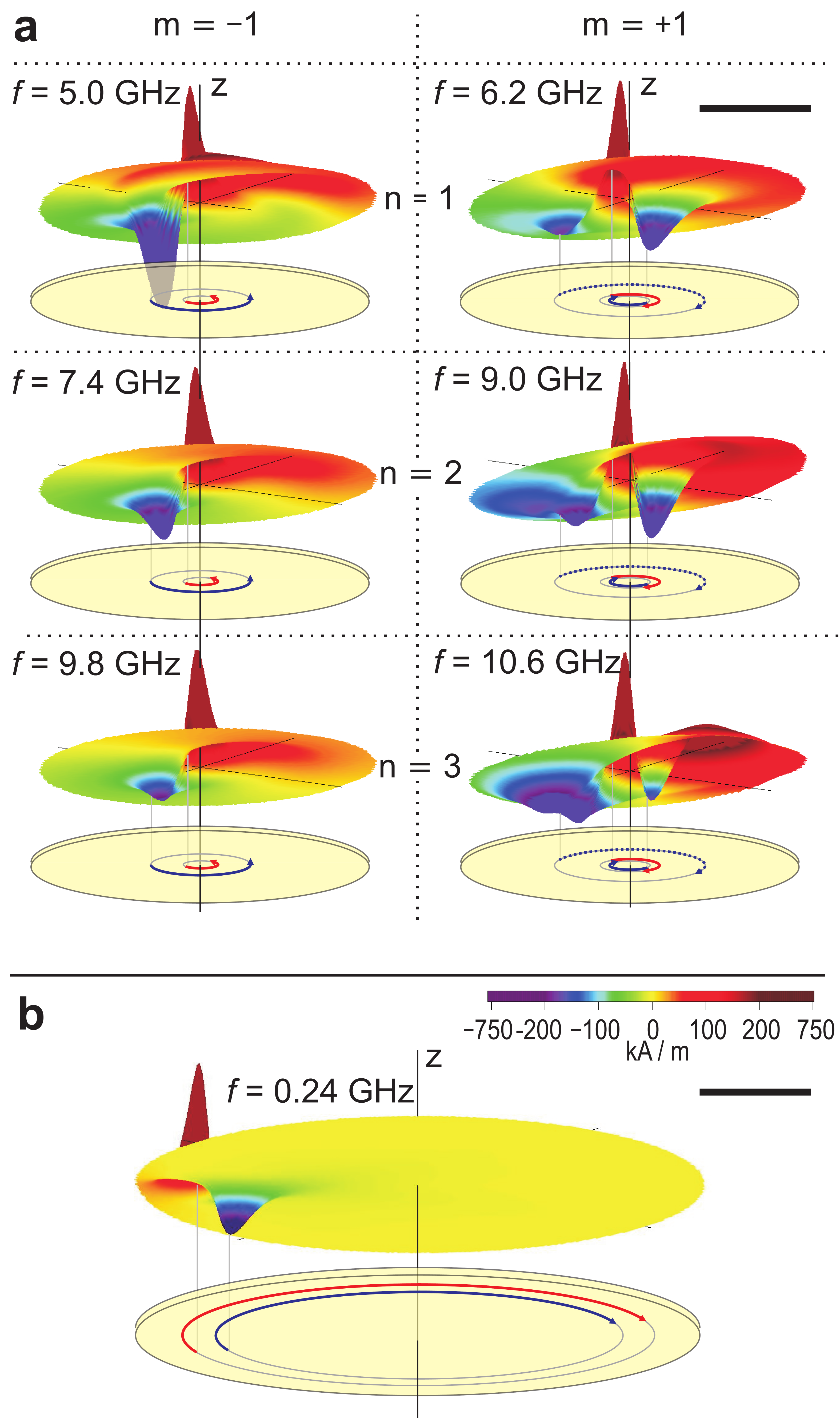}}
\caption{{\bf Snapshots of the 'dip' structures.} The out-of-plane magnetization of the central part extracted from the 1.6 $\mu$m disc before the core reversal. (a) A section of 300 nm in diameter showing the excited azimuthal spin wave modes as well as the 'dip'. The dynamics of modes with the same rotation sense are similar in the number of 'dips' and in their phase relation to the vortex core. This is indicated by the arrows below the snapshots. The difference between oppositely rotating modes are the result of the symmetry breaking due to the gyrofield. (b) For comparison the central part of the same structure with 500 nm in diameter is given, excited at the gyrotropic resonance. This results in a much larger trajectory of the core and the 'dip'.
The black bars correspond to a lateral size of 100 nm in both cases (a,b). 
}
\label{Fig3}
\end{figure}

\begin{figure}
\centerline{\includegraphics[clip, width=1.0\textwidth]{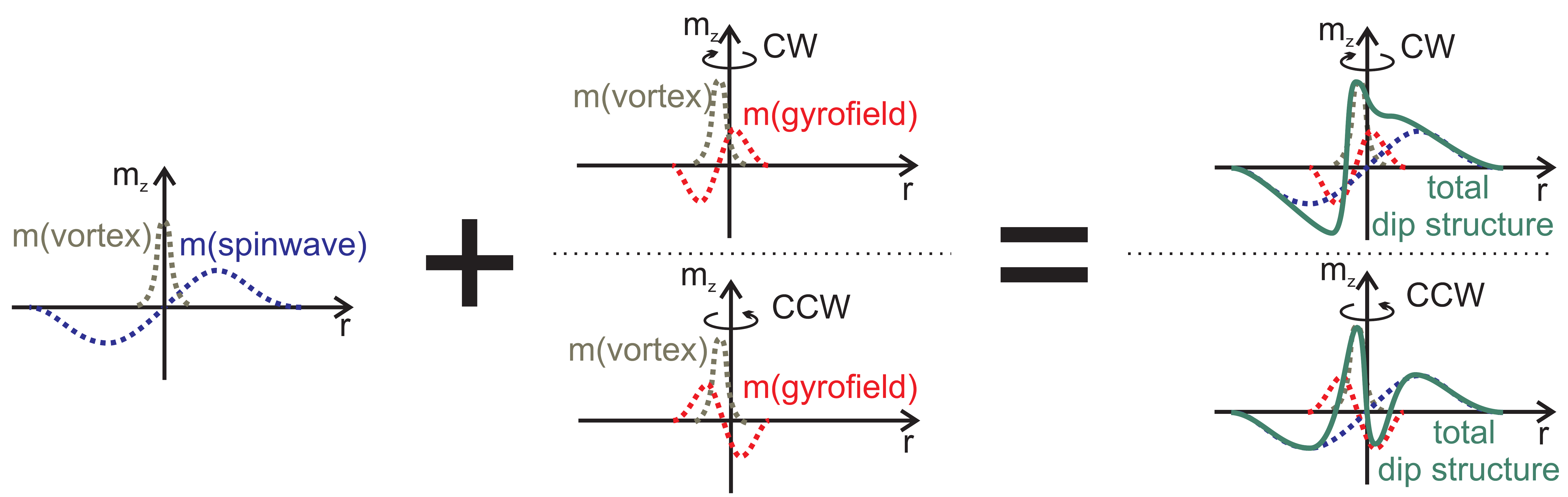}}
\caption{{\bf Model for spin wave induced vortex core reversal.} This sketch shows the origin of the out-of-plane magnetization near the vortex core and illustrates the origin of the observed asymmetry in the 'dip' formation for CW vs CCW spin wave excitation. To the left, the basic shape of the vortex core and the bipolar amplitude of the spin wave are shown. The magnetization change as a result of the gyrofield of the moving vortex core differs for CW and CCW rotation senses (middle sketches). The resulting structure (right sketches) agree qualitatively with results from the micromagnetic simulations as shown in Figs.~\ref{Fig3},~\ref{Fig5}~and~\ref{Fig7}.
}
\label{Fig8}
\end{figure}

The gyrofield, as used by Guslienko {\it et al.} to describe the 'dip' formation in the gyrotropic mode \cite{Thiele1973, Guslienko2008100}, can also be taken to explain the difference in 'dip' formation for the ($m=+1$) and ($m=-1$) modes, as is sketched in Fig.~\ref{Fig8}. This effective field originates from the movement of the vortex and results in out-of-plane contributions to the left and right side of the core (in respect to the direction of motion) which are respectively negative and positive (middle sketch in Fig.~\ref{Fig8}). This velocity dependent behaviour breaks the symmetry between counter rotating vortices. If we now combine the spin wave amplitude with the effect of the gyrofield, we can understand the 'dip' formation for any of the counter rotating modes (right sketch in Fig.~\ref{Fig8}). For a vortex up and the mode ($m=-1$) with a CW vortex motion, the gyrofield acts constructively on the spin wave amplitude. For the ($m=+1$) mode, it acts destructively, causing the double 'dip' configuration as seen in Fig.~\ref{Fig3}. Alternatively, these 'dip' structures can be regarded as a hybridisation between the gyrotropic mode and the azimuthal modes \cite{Guslienko2010}.
Kravchuk {\it et al.} \cite{Kravchuk2009} also presented a 'dip' formation mechanism for spin wave excitation under the condition of an artificially fixed vortex core.  However, our results show that this condition can not be used to predict the reversal in free vortices, as in realistic structures. According to their analysis, the polarization direction of the 'dip' is given by the rotation sense of the external field only. This would mean that reversal of a given core polarization is only possible for one of the two azimuthal modes. E.g. the reversal at the ($m=+1$) as presented in Fig.~\ref{Fig2} can not be explained with this assumption, as the 'dip' polarization would be the same as the core polarization. The dynamic origin of the 'dip' structure clearly explains this contradiction. For a free vortex, the 'dip' formation seems to be dominated by the gyrofield.

\begin{figure}
\centerline{\includegraphics[clip, width=0.5\textwidth]{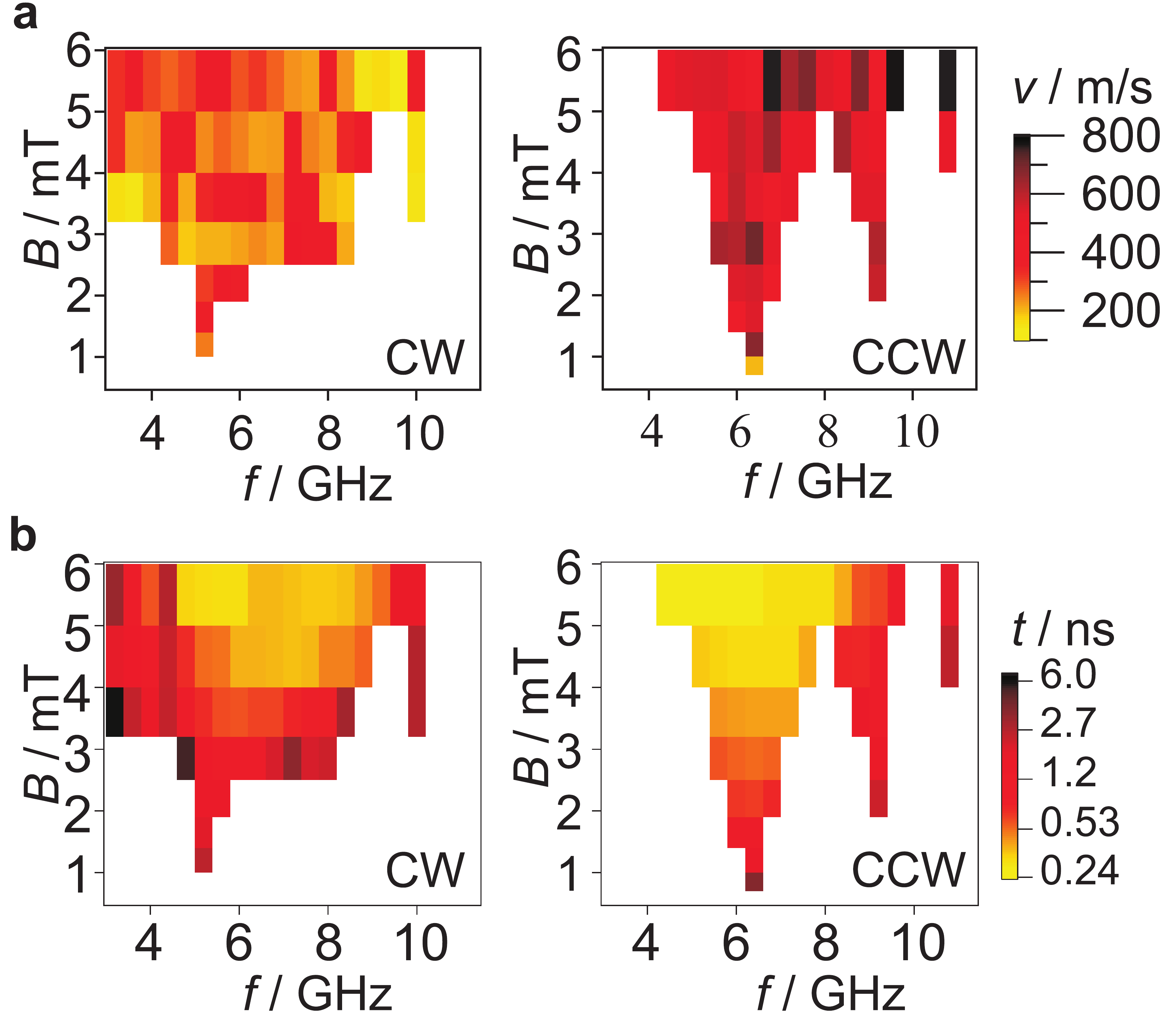}}
\caption{{\bf Vortex core velocities and switching times.} (a) Vortex core velocity just before switching. The gyrofield of the moving vortex is proportional to this velocity. Due to the important contribution of the spin wave background, this quantity is not a constant at GHz excitation and shows strong differences between CW and CCW excitation. (b) Excitation time until switching occurs in a logarithmic color scale. At sufficiently high amplitudes, switching takes in the order of one period of the excitation frequency, resulting in a widening of the resonances and a dominance of the mode ($n=1, m=+1$).}
\label{Fig5}
\end{figure}

Vortex core reversal by the low-frequency gyromode excitation results in a typical gyration radius of about 300 nm for this kind of sample. In contrast, the gyration radius of the vortex core at GHz frequencies is only in the order of 10 nm (cf. Fig.~\ref{Fig3}). But due to the high-frequency rotation, this small radius still results in a very high velocity of the vortex core. The top part of Fig.~\ref{Fig5} shows the velocities of the vortex core directly before the reversal process as calculated from the simulations. It varies between about 100 m/s ($m=-1$) and up to more than 600 m/s ($m=+1$). This demonstrates that for GHz spin wave excitation, no well-defined critical velocity is observed as it was calculated to be about 320 m/s for the gyromode of a Permalloy nanodot \cite{Guslienko2008100}.

The reason can be found in our explanation of the 'dip' formation as a combination of the gyrofield with the spin wave amplitude. Although in first order, the gyrofield is proportional to the vortex velocity, the spin wave amplitude is not and thus the critical velocity before switching is higher or lower than the expected value, depending if their combination is constructive or destructive. Similar arguments were used to explain the changes in critical velocity when perpendicular bias fields are applied \cite{Yoo2010}. The large fluctuations in the velocities can be understood when the random contribution of short wavelength spin waves are also taken into account. Consequently, the velocity itself is not a switching criterion for the spin wave mode induced vortex core reversal.

The bottom part of Fig.~\ref{Fig5} shows how long it takes for the vortex to switch after the burst is applied. It is no surprise that the larger the amplitude, the shorter the time for vortex core reversal. The minima in the excitation amplitude show a clear resonant behaviour. At high amplitudes the switching times become smaller than one period of the excitation and the frequency dependence is washed out, indicating the transition to non-resonant excitation \cite{Hertel2007}. However, the feature of unidirectionality gets lost in the intersection of two oppositely rotating modes if too short bursts are applied.

Because both the vortex and the created vortex-antivortex pair remain close to the centre of the structure during the excitation, the relaxation after the spin wave induced reversal process is much faster than in the gyrotropic case, where it has to spiral back from its orbit far away to the centre. In combination with the very small gyration radius, this fast relaxation is accelerating the whole reversal process by up to two orders of magnitude, while the total energy pumped into the system is roughly the same compared to reversal at the gyrotropic mode. This fact makes this new reversal process very beneficial for fast technological applications.

In conclusion, our experiments show vortex core reversal at GHz frequencies. By taking advantage of the vortex polarity dependent frequency splitting of the azimuthal modes ($n$, $m = \pm 1$) we present a concept for unidirectional reversal by exciting any of these modes up to 10 GHz.
Although the dynamics depend significantly on the different combinations of core polarity and azimuthal mode number, they can be explained by the superposition of the gyrofield with the spin wave amplitude. Consequently, immediately before the switching event, different vortex core velocities are observed, which excludes the velocity as a critical condition for core reversal at GHz excitation.
This finding highlights the importance of spin wave -- vortex interaction and boosts vortex core reversal to much higher frequencies, which may offer new routes for spintronics applications.

\section*{Methods}

\subsection*{Experiments}

Permalloy discs, 1.6\,$\mu$m in diameter and 50\,nm thickness, are grown on top of 2.5\,$\mu$m wide and 170\,nm thick copper striplines patterned on a 100\,nm thick silicon nitride membrane.
All structures are prepared by electron-beam lithography, thermal evaporation and lift-off processes.
AC currents with a phase shift of $\pm$90 degree flowing through orthogonal striplines are generating rotating in-plane magnetic fields \cite{Curcic2008}.
The out-of-plane magnetization of the Permalloy disc was imaged by scanning transmission X-ray microscopy at the MAXYMUS end station at HZB BESSY II, Berlin, by taking advantage of the X-Ray Magnetic Circular Dichroism (XMCD) \cite{SCHUTZ1987}. The instrument has a resolution of about 25\,nm, which is sufficient to resolve the vortex core and its polarization.

\subsection*{Simulations}

Micromagnetic simulations were performed with the Object Oriented Micromagnetic Framework (OOMMF) \cite{OOMMF}. A disc, 1.62\,$\mu m$ in diameter and 50\,nm thick, was simulated with a cell size of 3\,nm\,$\times$\,3\,nm\,$\times$\,50\,nm. Typical material parameters for Permalloy \cite{Weigand2009, Curcic2008} were used:
$M_s$=750\x10$^3$\,A/m, exchange constant $A$=13\x 10$^{-12}$\,J/m, damping parameter $\alpha$=0.006, anisotropy constant $K_1=0$.  For all simulations shown, the initial state was a vortex core up. The symmetry of the results were verified by also simulating with the opposite initial state which is a vortex down. The sample was excited by homogeneous in-plane rotating magnetic field bursts with duration of 24 periods.  The frequency of the bursts was scanned from 3\,GHz to 11\,GHz in 0.4\,GHz steps for both, CW and CCW sense of rotation.

\bibliographystyle{naturemag}

\begin{thebibliography}{10}
\expandafter\ifx\csname url\endcsname\relax
  \def\url#1{\texttt{#1}}\fi
\expandafter\ifx\csname urlprefix\endcsname\relax\def\urlprefix{URL }\fi
\providecommand{\bibinfo}[2]{#2}
\providecommand{\eprint}[2][]{\url{#2}}

\bibitem{Shinjo2000}
\bibinfo{author}{Shinjo, T.}, \bibinfo{author}{Okuno, T.},
  \bibinfo{author}{Hassdorf, R.}, \bibinfo{author}{Shigeto, K.} \&
  \bibinfo{author}{Ono, T.}
\newblock \bibinfo{title}{Magnetic vortex core observation in circular dots of
  permalloy}.
\newblock \emph{\bibinfo{journal}{Science}} \textbf{\bibinfo{volume}{289}},
  \bibinfo{pages}{930--932} (\bibinfo{year}{2000}).

\bibitem{Wachowiak2002}
\bibinfo{author}{Wachowiak, A.} \emph{et~al.}
\newblock \bibinfo{title}{Direct observation of internal spin structure of
  magnetic vortex cores}.
\newblock \emph{\bibinfo{journal}{Science}} \textbf{\bibinfo{volume}{298}},
  \bibinfo{pages}{577--580} (\bibinfo{year}{2002}).

\bibitem{Huber1982}
\bibinfo{author}{Huber, D.}
\newblock \bibinfo{title}{Equation of motion of a spin vortex in a
  two-dimensional planar magnet}.
\newblock \emph{\bibinfo{journal}{Journal of applied physics}}
  \textbf{\bibinfo{volume}{53}}, \bibinfo{pages}{1899--1900}
  (\bibinfo{year}{1982}).

\bibitem{Xiao2006}
\bibinfo{author}{Xiao, Q.}, \bibinfo{author}{Rudge, J.}, \bibinfo{author}{Choi,
  B.}, \bibinfo{author}{Hong, Y.} \& \bibinfo{author}{Donohoe, G.}
\newblock \bibinfo{title}{Dynamics of vortex core switching in ferromagnetic
  nanodisks}.
\newblock \emph{\bibinfo{journal}{Applied physics letters}}
  \textbf{\bibinfo{volume}{89}} (\bibinfo{year}{2006}).

\bibitem{Hertel2007}
\bibinfo{author}{Hertel, R.}, \bibinfo{author}{Gliga, S.},
  \bibinfo{author}{Fahnle, M.} \& \bibinfo{author}{Schneider, C.}
\newblock \bibinfo{title}{Ultrafast nanomagnetic toggle switching of vortex
  cores}.
\newblock \emph{\bibinfo{journal}{Physical Review Letters}}
  \textbf{\bibinfo{volume}{98}}, \bibinfo{pages}{117201}
  (\bibinfo{year}{2007}).

\bibitem{Weigand2009}
\bibinfo{author}{Weigand, M.} \emph{et~al.}
\newblock \bibinfo{title}{Vortex core switching by coherent excitation with
  single in-plane magnetic field pulses}.
\newblock \emph{\bibinfo{journal}{Physical Review Letters}}
  \textbf{\bibinfo{volume}{102}}, \bibinfo{pages}{077201}
  (\bibinfo{year}{2009}).

\bibitem{VanWaeyenberge2006}
\bibinfo{author}{{Van~Waeyenberge}, B.} \emph{et~al.}
\newblock \bibinfo{title}{Magnetic vortex core reversal by excitation with
  short bursts of an alternating field}.
\newblock \emph{\bibinfo{journal}{Nature}} \textbf{\bibinfo{volume}{444}},
  \bibinfo{pages}{461--464} (\bibinfo{year}{2006}).

\bibitem{Curcic2008}
\bibinfo{author}{Curcic, M.} \emph{et~al.}
\newblock \bibinfo{title}{Polarization selective magnetic vortex dynamics and
  core reversal in rotating magnetic fields}.
\newblock \emph{\bibinfo{journal}{Physical Review Letters}}
  \textbf{\bibinfo{volume}{101}}, \bibinfo{pages}{197204}
  (\bibinfo{year}{2008}).

\bibitem{Lee2008}
\bibinfo{author}{Lee, K.} \& \bibinfo{author}{Kim, S.}
\newblock \bibinfo{title}{Two circular-rotational eigenmodes and their giant
  resonance asymmetry in vortex gyrotropic motions in soft magnetic nanodots}.
\newblock \emph{\bibinfo{journal}{Physical Review B, Condensed matter and
  materials physics}} \textbf{\bibinfo{volume}{78}}, \bibinfo{pages}{014405}
  (\bibinfo{year}{2008}).

\bibitem{deLoubens2009}
\bibinfo{author}{{De Loubens}, G.} \emph{et~al.}
\newblock \bibinfo{title}{Bistability of vortex core dynamics in a single
  perpendicularly magnetized nanodisk}.
\newblock \emph{\bibinfo{journal}{Physical Review Letters}}
  \textbf{\bibinfo{volume}{102}}, \bibinfo{pages}{177602}
  (\bibinfo{year}{2009}).

\bibitem{Yamada2007}
\bibinfo{author}{Yamada, K.} \emph{et~al.}
\newblock \bibinfo{title}{Electrical switching of the vortex core in a magnetic
  disk}.
\newblock \emph{\bibinfo{journal}{Nature Materials}}
  \textbf{\bibinfo{volume}{6}}, \bibinfo{pages}{269--273}
  (\bibinfo{year}{2007}).

\bibitem{Yamada2008}
\bibinfo{author}{Yamada, K.}, \bibinfo{author}{Kasai, S.},
  \bibinfo{author}{Nakatani, Y.}, \bibinfo{author}{Kobayashi, K.} \&
  \bibinfo{author}{Ono, T.}
\newblock \bibinfo{title}{Switching magnetic vortex core by a single nanosecond
  current pulse}.
\newblock \emph{\bibinfo{journal}{Applied physics letters}}
  \textbf{\bibinfo{volume}{93}} (\bibinfo{year}{2008}).

\bibitem{Kim2008}
\bibinfo{author}{Kim, S.}, \bibinfo{author}{Lee, K.}, \bibinfo{author}{Yu, Y.}
  \& \bibinfo{author}{Choi, Y.}
\newblock \bibinfo{title}{Reliable low-power control of ultrafast vortex-core
  switching with the selectivity in an array of vortex states by in-plane
  circular-rotational magnetic fields and spin-polarized currents}.
\newblock \emph{\bibinfo{journal}{Applied Physics Letters}}
  \textbf{\bibinfo{volume}{92}}, \bibinfo{pages}{022509}
  (\bibinfo{year}{2008}).

\bibitem{Bohlens2008}
\bibinfo{author}{Bohlens, S.} \emph{et~al.}
\newblock \bibinfo{title}{Current controlled random-access memory based on
  magnetic vortex handedness}.
\newblock \emph{\bibinfo{journal}{Applied Physics Letters}}
  \textbf{\bibinfo{volume}{93}}, \bibinfo{pages}{142508}
  (\bibinfo{year}{2008}).

\bibitem{Pigeau2010}
\bibinfo{author}{Pigeau, B.} \emph{et~al.}
\newblock \bibinfo{title}{A frequency-controlled magnetic vortex memory}.
\newblock \emph{\bibinfo{journal}{Applied Physics Letters}}
  \textbf{\bibinfo{volume}{96}}, \bibinfo{pages}{132506}
  (\bibinfo{year}{2010}).

\bibitem{Guslienko2008100}
\bibinfo{author}{Guslienko, K.}, \bibinfo{author}{Lee, K.} \&
  \bibinfo{author}{Kim, S.}
\newblock \bibinfo{title}{Dynamic origin of vortex core switching in soft
  magnetic nanodots}.
\newblock \emph{\bibinfo{journal}{Physical Review Letters}}
  \textbf{\bibinfo{volume}{100}}, \bibinfo{pages}{027203}
  (\bibinfo{year}{2008}).

\bibitem{Vansteenkiste2009}
\bibinfo{author}{Vansteenkiste, A.} \emph{et~al.}
\newblock \bibinfo{title}{X-ray imaging of the dynamic magnetic vortex core
  deformation}.
\newblock \emph{\bibinfo{journal}{Nature Physics}}
  \textbf{\bibinfo{volume}{5}}, \bibinfo{pages}{332--334}
  (\bibinfo{year}{2009}).

\bibitem{Thiele1973}
\bibinfo{author}{Thiele, A.~A.}
\newblock \bibinfo{title}{Steady-state motion of magnetic domains}.
\newblock \emph{\bibinfo{journal}{Phys. Rev. Lett.}}
  \textbf{\bibinfo{volume}{30}}, \bibinfo{pages}{230--233}
  (\bibinfo{year}{1973}).

\bibitem{Choe2004}
\bibinfo{author}{Choe, S.~B.} \emph{et~al.}
\newblock \bibinfo{title}{Vortex core-driven magnetization dynamics}.
\newblock \emph{\bibinfo{journal}{Science}} \textbf{\bibinfo{volume}{304}},
  \bibinfo{pages}{420--422} (\bibinfo{year}{2004}).

\bibitem{Kravchuk2007}
\bibinfo{author}{Kravchuk, V.}, \bibinfo{author}{Sheka, D.},
  \bibinfo{author}{Gaididei, Y.} \& \bibinfo{author}{Mertens, F.}
\newblock \bibinfo{title}{Controlled vortex core switching in a magnetic
  nanodisk by a rotating field}.
\newblock \emph{\bibinfo{journal}{Journal of Applied Physics}}
  \textbf{\bibinfo{volume}{102}}, \bibinfo{pages}{043908}
  (\bibinfo{year}{2007}).

\bibitem{Kravchuk2009}
\bibinfo{author}{Kravchuk, V.}, \bibinfo{author}{Gaididei, Y.} \&
  \bibinfo{author}{Sheka, D.}
\newblock \bibinfo{title}{Nucleation of a vortex-antivortex pair in the
  presence of an immobile magnetic vortex}.
\newblock \emph{\bibinfo{journal}{Physical Review B, Condensed matter and
  materials physics}} \textbf{\bibinfo{volume}{80}}, \bibinfo{pages}{100405}
  (\bibinfo{year}{2009}).

\bibitem{Park2003}
\bibinfo{author}{Park, J.}, \bibinfo{author}{Eames, P.},
  \bibinfo{author}{Engebretson, D.}, \bibinfo{author}{Berezovsky, J.} \&
  \bibinfo{author}{Crowell, P.}
\newblock \bibinfo{title}{Imaging of spin dynamics in closure domain and vortex
  structures}.
\newblock \emph{\bibinfo{journal}{Physical Review B, Condensed matter and
  materials physics}} \textbf{\bibinfo{volume}{67}}, \bibinfo{pages}{020403}
  (\bibinfo{year}{2003}).

\bibitem{Buess2005}
\bibinfo{author}{Buess, M.} \emph{et~al.}
\newblock \bibinfo{title}{Excitations with negative dispersion in a spin
  vortex}.
\newblock \emph{\bibinfo{journal}{Physical Review B, Condensed matter}}
  \textbf{\bibinfo{volume}{71}}, \bibinfo{pages}{104415}
  (\bibinfo{year}{2005}).

\bibitem{Buess2004}
\bibinfo{author}{Buess, M.} \emph{et~al.}
\newblock \bibinfo{title}{Fourier transform imaging of spin vortex eigenmodes}.
\newblock \emph{\bibinfo{journal}{Physical Review Letters}}
  \textbf{\bibinfo{volume}{93}}, \bibinfo{pages}{077207}
  (\bibinfo{year}{2004}).

\bibitem{neudecker2006}
\bibinfo{author}{Neudecker, I.} \emph{et~al.}
\newblock \bibinfo{title}{Modal spectrum of permalloy disks excited by in-plane
  magnetic fields}.
\newblock \emph{\bibinfo{journal}{Physical Review B}}
  \textbf{\bibinfo{volume}{73}}, \bibinfo{pages}{134426}
  (\bibinfo{year}{2006}).

\bibitem{Zaspel2005}
\bibinfo{author}{Zaspel, C.}, \bibinfo{author}{Ivanov, B.}, \bibinfo{author}{J}
  \& \bibinfo{author}{Crowell, P.}
\newblock \bibinfo{title}{Excitations in vortex-state permalloy dots}.
\newblock \emph{\bibinfo{journal}{Physical Review B, Condensed matter and
  materials physics}} \textbf{\bibinfo{volume}{72}}, \bibinfo{pages}{024427}
  (\bibinfo{year}{2005}).

\bibitem{Park2005}
\bibinfo{author}{Park, J.} \& \bibinfo{author}{Crowell, P.}
\newblock \bibinfo{title}{Interactions of spin waves with a magnetic vortex}.
\newblock \emph{\bibinfo{journal}{Physical Review Letters}}
  \textbf{\bibinfo{volume}{95}}, \bibinfo{pages}{167201}
  (\bibinfo{year}{2005}).

\bibitem{Zhu2005}
\bibinfo{author}{Zhu, X.}, \bibinfo{author}{Liu, Z.},
  \bibinfo{author}{Metlushko, V.}, \bibinfo{author}{Grutter, P.} \&
  \bibinfo{author}{Freeman, M.}
\newblock \bibinfo{title}{Broadband spin dynamics of the magnetic vortex state:
  Effect of the pulsed field direction}.
\newblock \emph{\bibinfo{journal}{Physical Review B, Condensed matter and
  materials physics}} \textbf{\bibinfo{volume}{71}}, \bibinfo{pages}{180408}
  (\bibinfo{year}{2005}).

\bibitem{Hoffmann2007}
\bibinfo{author}{Hoffmann, F.} \emph{et~al.}
\newblock \bibinfo{title}{Mode degeneracy due to vortex core removal in
  magnetic disks}.
\newblock \emph{\bibinfo{journal}{Physical Review B, Condensed matter and
  materials physics}} \textbf{\bibinfo{volume}{76}} (\bibinfo{year}{2007}).

\bibitem{Guslienko2008101}
\bibinfo{author}{Guslienko, K.~Y.}, \bibinfo{author}{Slavin, A.~N.},
  \bibinfo{author}{Tiberkevich, V.} \& \bibinfo{author}{Kim, S.-K.}
\newblock \bibinfo{title}{Dynamic origin of azimuthal modes splitting in
  vortex-state magnetic dots}.
\newblock \emph{\bibinfo{journal}{Physical Review Letters}}
  \textbf{\bibinfo{volume}{101}}, \bibinfo{pages}{247203}
  (\bibinfo{year}{2008}).

\bibitem{Guslienko2010}
\bibinfo{author}{Guslienko, K.}, \bibinfo{author}{Aranda, G.} \&
  \bibinfo{author}{Gonzalez, J.}
\newblock \bibinfo{title}{Topological gauge field in nanomagnets: Spin-wave
  excitations over a slowly moving magnetization background}.
\newblock \emph{\bibinfo{journal}{Physical Review B, Condensed matter and
  materials physics}} \textbf{\bibinfo{volume}{81}}, \bibinfo{pages}{014414}
  (\bibinfo{year}{2010}).

\bibitem{OOMMF}
\bibinfo{author}{Donahue, M.} \& \bibinfo{author}{Porter, D.}
\newblock \bibinfo{title}{Oommf user's guide, version 1.0}.
\newblock \emph{\bibinfo{journal}{Interagency Report NISTIR 6376, National
  Institute of Standards and Technology, Gaithersburg, MD}}
  (\bibinfo{year}{1999}).

\bibitem{Volkel1994}
\bibinfo{author}{V\"olkel, A.~R.}, \bibinfo{author}{Wysin, G.~M.},
  \bibinfo{author}{Mertens, F.~G.}, \bibinfo{author}{Bishop, A.~R.} \&
  \bibinfo{author}{Schnitzer, H.~J.}
\newblock \bibinfo{title}{Collective-variable approach to the dynamics of
  nonlinear magnetic excitations with application to vortices}.
\newblock \emph{\bibinfo{journal}{Phys. Rev. B}} \textbf{\bibinfo{volume}{50}},
  \bibinfo{pages}{12711--12720} (\bibinfo{year}{1994}).

\bibitem{Yoo2010}
\bibinfo{author}{Yoo, M.-W.}, \bibinfo{author}{Lee, K.-S.},
  \bibinfo{author}{Jeong, D.-E.} \& \bibinfo{author}{Kim, S.-K.}
\newblock \bibinfo{title}{Origin, criterion, and mechanism of vortex-core
  reversals in soft magnetic nanodisks under perpendicular bias fields}.
\newblock \emph{\bibinfo{journal}{Phys. Rev. B}} \textbf{\bibinfo{volume}{82}},
  \bibinfo{pages}{174437} (\bibinfo{year}{2010}).

\bibitem{SCHUTZ1987}
\bibinfo{author}{Schutz, G.} \emph{et~al.}
\newblock \bibinfo{title}{Absorption of circularly polarized x-rays in iron}.
\newblock \emph{\bibinfo{journal}{Physical Review Letters}}
  \textbf{\bibinfo{volume}{58}}, \bibinfo{pages}{737--740}
  (\bibinfo{year}{1987}).

\end{thebibliography}

\begin{acknowledgments}
We would like to thank Manfred Fähnle, MPI Stuttgart, Franz Mertens, Bayreuth University, Yuri Gaididei and Volodymyr Kravchuk, Bogolyubov Institute for Theoretical Physics, Kiev, and Denis Sheka, Kiev University, for fruitful discussions.  We would also like to thank all people involved in the construction and operation of the MAXYMUS scanning X-ray microscope at HZB BESSY II in Berlin, in particular Brigitte Baretzky, Michael Bechtel, Rolf Follath, Corinne Gr{\'e}vent and Eberhard Goering. Cooperation with Rolf Heidemann, Ulm University, in the stripline design for microwave application is gratefully acknowledged.
\end{acknowledgments}

\subsection*{Author contributions}
Project planning: M.K., H.S., B.V.W., G.S.; Sample preparation: G.W., C.H.B.; Experiments: M.K., M.N., M.S., M.W., M.C.; Micromagnetic simulations: M.K.; Analyse data: M.K., M.C., A.V., B.V.W., H.S.; Writing the paper: M.K., M.N., M.S., M.C., M.W., A.V., B.V.W., H.S., G.W., C.H.B., G.S.

\subsection*{Competing financial interests}
The authors declare that they have no competing interests as defined by Nature Publishing Group, or other interests that might be perceived to influence the results and/or discussion reported in this paper.

\newpage


\end{document}